\documentclass[a4paper
]{article}
\usepackage{amssymb}
\usepackage{amsmath}
\usepackage{latexsym}

\def\lr{Lachi\`eze-Rey} 

\def\R{{\rm I\!R}}
\def\bbbC {\mathbb{C}}
\def\setC {\mathbb{C}}

\def\Tr {\rm Tr}

\def\tendsto {\rightarrow}
\def\on {orthonormal}
\def\nc {non commutative}

\def \I {\mbox{1\hspace{-.35em}1}}
\def\calH {{\cal H}}
\def\ie {{i.e.}}

\def\guill {\textquotedblleft}

\newcommand{\norm}[1]{\mid #1 \mid}
\newcommand{\ket}[1]{\mid #1 \,\rangle}
\newcommand{\bra}[1]{\langle \,#1 \mid}
\newcommand{\braket}[2]{\langle \, #1 \mid #2 \,\rangle}
\newcommand{\matrixdd}[4]{\left[\begin{array}{cc}#1&#2\\#3&#4\\ 
\end{array} \right]}

\newcommand{\Id}[1]{1\hspace{-.35em}1_{#1}}

\begin{document} 
\title{Quantization of the sphere  with coherent states}

\author{preprint{APS/CS}\\ 
Marc Lachi\`eze Rey$^{1,2} $,\\
Jean-Pierre Gazeau$^{1,3} $, Eric Huguet$^{1,4} $, 
 Jacques  Renaud$^{1,5} $\\ and  Tarik Garidi$^{1,6} $,\\
1- F\'ed\'eration de Recherches 
Astroparticules et Cosmologie\\ Boite 7020, Universit\'e Paris 7  
Denis Diderot,\\
F-75251 Paris Cedex 05, France \\
2~Service d'Astrophysique, C.E. Saclay,\\ 91191 Gif sur Yvette cedex, 
France, marclr@cea.fr\\
3~gazeau@ccr.jussieu.fr\\
4~eric.huguet@obspm.fr\\
5~renaud@ccr.jussieu.fr
6~garidi@ccr.jussieu.fr
} 
 
\maketitle
 \section{Main ideas}

Current views link   quantization with dynamics. The reason is that   
quantum mechanics or quantum field theories address to dynamical 
systems, \ie, particles or fields. Our  point of view    here 
  breaks the link between quantization and dynamics: any (classical) 
physical system can be quantized. Only    dynamical systems lead to   
\emph{dynamical} quantum theories, which appear to      result from   
the quantization of  symplectic structures.

The procedure developed here,  through \emph{coherent states} (CS),  
allows to   quantize   any system considered as  an {\sl Observation 
Set}, \ie, a set of data  $X = \{x\}$, whose  elements  can be 
points, or any kind   of parameters. When $X$ has a symplectic 
structure, it can be considered as   a  phase space, and   our  
approach is then  equivalent to the usual quantization, although with 
some peculiar characteristics. But the CS procedure is much more 
general and can be applied  even in the absence of   symplectic 
structure, and in fact of any   structure at all  (other than a 
measure) over $X$.

A  quantization, in this sense,  may   be  considered as a different 
way to look at the system. It shows numerous analogies    with some  
procedures used in  data handling (discussed in more details in 
Gazeau et al. 2003 \cite{gghlr}), for instance those involving  
wavelets, which are the basic  example of   coherent states. In many 
respects, the   choice of a quantization   appears here as the  
choice of a resolution to look at the  system. 
As it is well known, some aspects of  quantum mechanics may be seen  
as a \nc ~geometry of the phase space (position and momentum 
operators do not commute). If we quantize a \guill ~space of data ", 
it will be no surprising that a \nc  ~geometry emerges. We will show 
explicitly how a quantization of the ordinary sphere leads    to its  
fuzzy  geometry.

 \section{Coherent states}
 
The      (classical) system to be quantized is considered as   a set 
of data,  $X = \{x\}$, with no other   specified   structure than   a 
measure $\mu$ (with measure axioms; see   \cite{gghlr,Elliott}). 
The quantization is defined by the choice of  a (separable) Hilbert 
space  $\calH$ with    an inclusion  map 
\begin{equation} \label{map}
X \ni x\, \mapsto \, \ket{x}~\in \calH.
\end{equation} This defines   the coherent states $\ket{x}$, which  
must  obey two conditions: \\
- the resolution of the identity:\begin{equation} \label{overcomp}
\int_X \mu(dx) \, \ket{x} \bra{x} \, =\, \Id{\calH}.\end{equation} It 
 implies that the {\sl coherent states} ${\displaystyle \ket{x}}$ 
form an
{\it over-complete} (continuous) basis of $\calH$.\\
- a normalization
\begin{equation}\label{norm}
\braket{x}{x}\,=\,1.
\end{equation}
 Note that the  $\ket{x} \bra{x}$ appear as natural Hermitian   
operators (orthogonal projection)  over ${\calH}$.

There  exists  a natural Hilbert space associated to   $X, \mu$: the 
space $L^{2}(X,\mu)$ of square-integrable functions over $X$. There 
is an isometric embedding $W$ of our (closed) Hilbert  subspace  
$\calH$   in  $L^{2}$,   resulting from   the \emph{Weyl-Wigner 
injection} \begin{equation}
{\displaystyle  \calH \ni   \,\ket{\psi} \mapsto \Psi \in W({\cal
    H}) \subset L^{2} :~x \mapsto  \Psi(x)\,\equiv \,\braket{x}{\psi} 
} .\end{equation} 
Thus, the quantization procedure   may also be seen  as a peculiar 
choice of a sub-space of~$L^{2}$. 
An explicit procedure   is explicited in \cite{gghlr}, and applied 
below to the sphere $S^2$.
It begins with  the selection of  a sub-vector space of  $L^2$ by 
defining an \on ~set of $N$ functions $\phi _{i} $, verifying   
${\cal N} (x) \equiv \sum _{i=1}^{N} \mid  \phi  _{i} (x) \mid ^2 < 
\infty$.
We note $\phi _{i} $, as a vector,  with the ket notation $\ket{i}$, 
and we define the family of coherent states as \begin{equation}
\ket{x} \equiv \frac{1}{{\cal N} (x)} \sum _i \phi _i 
(x)~\ket{i},\end{equation}
which allows to  perform the analysis presented above.
The resolution of identity implies the existence of a {\it 
reproducing kernel}~$K$, in $\calH$ considered as a subset of  
$L^{2}$,  such that
$\Psi = K \circ \Psi$ (cf. wavelets), \ie,
\begin{equation}
\Psi(x)\,=\,\int \mu(dy)~K(x,y)~\Psi(y);\ ~~~~K(x,y) 
\,\equiv\,\braket{x}{y}.\end{equation}

\subsection{Observables  and Symbolic calculus}

A \emph{classical}  observable over $X$ is a  function $f:X \mapsto 
K$ ($\R$ or $\bbbC$). 
To any such function $f$, we associate the    observable  over 
${\calH}$,   \begin{equation}
A_{f} \equiv \int_X~ \mu(dx)~f(x)~  \, \ket{x} \bra{x} .
\end{equation} For a large class of observables, these operators are  
self-adjoint.

The existence of the  continuous frame $\{\ket{x}\}$ enables the 
definition of a symbolic calculus {\it \`a la} Berezin-Lieb 
\cite{ber}. To each  linear, self-adjoint operator (observable) 
${\cal O}$ acting on 
$\calH$, one associates the {\sl   lower (or covariant) symbol}
\begin{equation} \label{lsymb}
\check{\cal O}(x)\,\equiv \,\bra{x} {\cal O} \ket{x},
\end{equation}    
and the {\sl  upper (or contravariant) symbol} (not necessarily 
unique) $\hat{\cal O}$ such that
\begin{equation}\label{usymb} {\cal O} \,=\,\int_X \mu(dx) ~\hat{\cal 
O}(x) \ket{x} \bra{x}.
\end{equation}
Note that $f$ is an upper symbol of $A_{f}$.

They obey the  Berezin-Lieb  inequalities: \begin{equation}
\int_X \mu(dx)\, g \bigl(\check{\cal O}(x)\bigr) \,  \leqslant  \Tr\, 
g\bigl({\cal O}\bigr) \,
\leqslant \int_X \mu(dx)\, g\bigl(\hat {\cal O}(x)\bigr),
\end{equation}
\noindent where $g(x)$ is a convex map.

\subsection{First example: quantization of the circle  $  {\cal S}^1$}

In \cite{gghlr},  we gave the simplest examples of application of 
this procedure: the quantization of a discrete set of elements, and 
of the unit interval.  Here we follow by showing a   quantization 
of    the circle $X=\{\theta\}$, with the normalized measure $d\theta 
/\pi$.

The simplest possibility is  a (real) quantization, with  ${\cal H} = 
\R 
^2:=\{(x,y)\}$.  The map (\ref{map}) is  defined by \begin{equation}
X \ni \theta \mapsto \ket{\theta} \equiv (\cos \theta, \sin \theta) 
\in \R ^2.
\end{equation}
 
The coherent states  $\ket{\theta}$ are the unit vectors of $\R^2$ of 
argument 
$\theta$, which  design the unit circle (thus, the embedding).
 It is easy to check that they  form an over-complete basis of 
$\calH$, with the   completness
relation (\ref{overcomp}): \begin{equation} 
\int_0^{2 \pi} \frac{d\theta}{\pi} \, \ket{\theta} \bra{\theta} \, 
=\, \I _{\calH}.
\nonumber \end{equation}

In matrix notation, an observable  of $\R^2$ is written as a linear 
symetric $2\times2$ 
matrix  \begin{equation}
A\,=\,\matrixdd{a}{b}{b}{d}=\frac{a + d}{2}~\sigma _0+\frac{a - 
d}{2}~ 
\sigma _{3}+b~\sigma _{1}, 
\end{equation}  so that $\sigma _0 \equiv \I$ and the Pauli matrices 
$\sigma _ i$  form  a basis for 
the space of observables. 

Corresponding upper and lower symbols can be  obtained as 
\begin{equation}
\check{A}(\theta)\,=\,\frac{a + d}{2} \,+\, \frac{a - d}{2} 
\cos2\theta
 \,+\, b~ \sin2\theta , \nonumber
\end{equation}
\begin{equation}
\hat{A}(\theta)\,=\,\frac{a + d}{2} \,+\, (a - d) ~\cos2\theta
 \,+\, b ~\sin2\theta . \nonumber \end{equation}

\section{Quantizations  of the 2-sphere}

\subsection{The 2-sphere}

In \cite{gghlr}, we  proposed a practical method to construct 
explicitly the coherent states by selecting some peculiar elements of 
$L^2$.  Here we   apply this method to the quantization of   the  
Observation Set $X =S^{2}$, the 2-sphere. A  point of $X$ is noted  
$x = (\theta,\phi)$. We adopt the normalized measure $\mu (dx) = \sin 
\theta ~d\theta~d\phi/4\pi$, proportional to  the
$SU(2)-$invariant   measure, which is also the surface element. 

We know that $\mu$   is
a symplectic form,  with the canonical coordinates $q=\phi,\,
p=-\cos \theta$. This allows to see  $S^{2}$  as the   phase space 
for the    theory of (classical) angular momentum. In this spirit, we 
will be able to    interpret our  procedure as the construction of 
the coherent spin states. Also, our construction will take advantage 
of the group action of SO(3) on $S^2$. 
$S^{2}$ is embedded in $\R ^{3}$, and   $G=SO(3)$ acts  as  isometry  
group in $S^{2}$.  However, we emphasize again that our quantization 
procedure is based on the only existence of a   measure,    and may 
be used in the absence of   metric or symplectic structure.

Quantization is defined by  an embedding of $S^{2}$ in an  Hilbert 
$\calH$.  This paper deals with  the   simple case  $\calH=\bbbC 
^{2}$. The cases  $\calH=  \R ^{n}$ will be treated in future works.

\subsection{Two-dimensional hermitian processing of the 2-sphere: the
quantum spin in its complex version}

Here, we  embed $S ^2$ into     the smallest complex  Hilbert space 
possible $\calH = 
\setC^{2}$. This  quantization leads to the {\sl  coherent 
spin states} \cite{Perelomov,Radcliffe,Arecchi}.
 Proceeding as indicated in \cite{gghlr},  we define $\calH$ by the 
selection of  the two complex functions
\begin{eqnarray}\label{fud}
\ket{1} \equiv \Phi_{1}: \Phi_{1}(x)&=&\sqrt{2}~\cos\theta/2  \\
\ket{2} \equiv \Phi_{2}: \Phi_{2}(x)&=&\sqrt{2}~\sin\theta/2 
~e^{i\phi},\hspace{2cm}
0\leq\theta\leq\pi\;,\quad 0\leq\phi\leq2\pi\;.\nonumber
\end{eqnarray}

We define the embedding map  \begin{equation} x \mapsto \ket{x} 
\,=\sqrt{2}~\cos \theta /2
\,\ket{1} \,+\,\sqrt{2}~ \sin  \theta /2 \,e^{i\,\phi} 
\,\ket{2},\end{equation}
leading to   \begin{eqnarray} \ket{x} \bra{x}
&=&2~ \matrixdd{\cos ^{2 }\theta/2}{\cos  \theta/2~\sin
\theta/2~e^{-i~\phi} }{\cos \theta/2~\sin \theta/2~e^{ i~\phi}
}{\sin ^{2 }\theta/2}  
\nonumber\\&=&[\sigma_{0}+\cos
\theta ~\sigma_{3}+ \sin\theta ~\cos \phi ~\sigma_{1} + \sin\theta
~\sin \phi ~\sigma_{2}].\nonumber\end{eqnarray}

We can check
\begin{equation}\int _{S^{2} } \mu(dx) \,\ket{x} 
\bra{x}\,=\,\I,\hspace{2cm}
\braket{x}{x}\,=\,1,\nonumber
\end{equation}
  
The Pauli matrices $\sigma_{i}$ and $ \sigma_{0}$ form a basis of the
$2\times 2$ complex hermitian matrices. The upper and lower symbols 
follow from those of the basis, namely
\begin{eqnarray*}
\check{\sigma}_0 &=& 1\\
\check{\sigma}_1 &=& \sin\theta\,\cos\phi \\
\check{\sigma}_2 &=& \sin\theta\,\sin\phi \\
\check{\sigma}_3 &=& \cos\theta \\
\hat{\sigma}_0 &=& 1 \\
\hat{\sigma}_i &=& 3\, \check{\sigma}_i ,\;i = 1,2,3.
\end{eqnarray*}

We obtain easily the operators associated to the functions
(coordinates) $\theta$ and $\phi$ as \begin{equation}
A_{\theta}=\frac{\pi}{8}~\matrixdd{3}{0}{0}{5},
A_{\phi}=\frac{\pi}{4}~\matrixdd{4}{i}{-i}{4}. \end{equation}
Their commutator is $
[A_{\phi},A_{\theta}]=\frac{i~\pi ^{2}}
{64}~\sigma _1$, with
\begin{equation}
\bra{x}  [A_{\phi},A_{\theta}] ~\ket{x}= \frac{i~\pi^2}{64}~ 
\sin(\theta)~\cos(\phi)
\end{equation} and
\begin{equation}
\bra{x}  [A_{\phi},A_{\theta}]^{2} ~\ket{x}=
-\frac{\pi^{4}}{(64)^2}.
\end{equation}

We may calculate the operators associated to the coordinates in $\R 
^3$:
\begin{eqnarray}
x^1=\sin \theta~\cos\varphi & \mapsto  & \frac{1}{3}~\sigma _1  \\
x^2=\sin \theta~\sin \varphi & \mapsto  & \frac{1}{3}~\sigma _2  \\
x^3=\cos \theta  & \mapsto  & \frac{1}{3}~\sigma _3,
\end{eqnarray} involving the three Pauli matrices. These operators 
provide the quantum version of the coordinates. We interpret them 
below in terms of \nc ~geometry.

Note that we can perform the same   procedure with the two functions 
$  \Phi '_{1}(x)=\sqrt{2}~\cos\theta/2~e^{-i\phi/2}  $ and  $  \Phi ' 
_{2}(x)=\sqrt{2}~\sin\theta/2 ~e^{i\phi/2}$, instead of $  \Phi _{1}$ 
and $  \Phi _{2}$ given by   (\ref{fud}), with identical results.

The generalization to $L+1$, instead of 2   dimensions   starts from  
a choice of $L+1$  basis functions (see below), leading to  an 
Hilbert space $\calH$ of dimension $L+1$. As we will see, this is 
linked to the fuzzy sphere with  $L+1$ cells.

\subsection{Link with the fuzzy sphere}

We recall an usual construction of the fuzzy sphere  (\cite{Madore} 
p.148).
It starts from the decomposition of any  smooth function $f \in 
C(S^2)$ in spherical harmonics,\begin{equation} \label{develop}
f=\sum _{{\ell}=0}^{\infty}~\sum _m~f_{{\ell}m}~Y^{\ell}_m.
\end{equation} We note $V^{\ell }$ the   $(2 \ell +1)$-dimensional 
vector space generated by the  $Y^{\ell}_m$, for fixed~$\ell$. The 
direct sum  $\bigoplus _{\ell=0}^L V^{\ell }$, generated by the  
$Y^{\ell}_m$ for $\ell \le L$, is a vector space of dimension    
$(L+1)^2$.

Through the  embedding of  $S^2$ in $\R ^3$,  we may  write each  
point of  $ S^2$ as $x=(x^i)$, with $\sum _{i=1}^3(x^i)^2=1$.
Any function in $S^2$ can be seen as the restriction of a function 
on~$\R ^3$. Moreover, such functions are generated by the homogeneous 
polynomia in $\R ^3$.
This allows (identifying a function and its restriction) to  express 
equation (\ref{develop})  in  a  polynomial form in $\R 
^3$:\begin{equation}
\label{developPol}
f=f_{(0)}+\sum _{(i_1)}~f_{(i)}~x^i+...+\sum _{(i_1 i_2 ...  
i_{\ell})} ~ 
f_{(i_1i_2...i_{\ell})}~x^{i_1}~x^{i_2}...x^{i_{\ell}}+...,
\end{equation}
where each sum extends  to all symmetric combinations of the $\ell$  
indices to generate $V^{\ell}$. For each fixed value of $\ell$, the 
$2{\ell}+1$  coefficients $f_{(i_1i_2...i_{\ell})}$ ($\ell$ fixed) 
are those of the symmetric traceless 
$3 \times 3 \times ...\times 3$ (${\ell}$ times) matrices.

To obtain $S_{fuzzy,L+1}$,  the fuzzy sphere with $L+1$ cells,\\
- we consider  the three  generators $J^i$ of the $L+1$ dimensional 
irreducible unitary representation (IUR) of SU(2). They are  
expressed as   $(L+1) \times (L+1) $ matrices obeying
\begin{equation}
[J_i,J_J]=i~\epsilon _{ijk}~J_k.
\end{equation}
- To obtain the  operator $F$ associated to  any function  $f$, we 
first replace each  $x^i$ by the matrix $X^i=\kappa~J^i$, where  
$\kappa=2 r/\sqrt{L^2+2L}$.\\
- In  the above development     (\ref{developPol}) of  $f$, 
we replace each coordinate $x^i$  by   the $(L+1) \times (L+1) $ 
matrix $X^i$, and the usual product by matrix  product. Then  we 
truncate the expression obtained at index $\ell=L$. These matrices  
generate the   set $M^{L+1}$ of   $(L+1)^2$ independent   $(L+1) 
\times (L+1) $ matrices: a closed algebra  through the product, 
which  provides  a finite approximation to $C(S^2)$.  According to 
this construction, a basis of $M^{L+1}$ is provided by all  
products of the $J^i$'s up to power~$L$.
The corresponding  (\nc) matrix geometries  are  finite, fuzzy 
approximations to  the smooth sphere  $S^2$, which appears as the 
limit $N \tendsto \infty$ of their  sequence. Note that $M^{L+1}$ may 
be identified to
$\bigoplus _{\ell=0}^L V^{\ell }$.

Examples: \begin{itemize}
  \item  L=0: we replace the $x^i$ by the pure number $1$ and $M^1$, 
of dimension 1, reduces   to $\bbbC$.
  \item   L=1:  we replace the $x^i$ by $\kappa _1  \sigma ^i$, the 
three Pauli matrices  ($\kappa _1=2r/3$).  By their products, they 
generate $M^2$, of dimension 4. This gives the geometry of the fuzzy 
sphere $S_{fuzzy,2}$  with 2 cells.
  \item 
 L=2: we replace the $x^i$ by $\kappa _2  J^i$, with $\kappa 
_2=r/\sqrt{2}$, and  the three rotation matrices; 
$[J^i,J^j]=i~\epsilon _{ijk}~J^{k}$. By their products, they generate 
$M^3$, of dimension 9. This gives the geometry of the fuzzy sphere 
$S_{fuzzy,3}$ with 3 cells.
\end{itemize}

According to this construction, the geometry of the fuzzy sphere 
results from the choice of the algebra $M^{L+1}$, of the  
representation  matrices,  with their matrix  product. This gives  
the  abstract algebra of operators acting on $S_{fuzzy,L+1}$. 
The  order    $(L+1)$ of the matrices invites to see them as acting 
as the  endomorphisms of  an Hilbert space of dimension $(L+1)$ 
\cite{Freidel}.  This is exactly what provides the coherent states 
introduced here.

{\bf Fuzzy spheres from coherent states}

The CS procedure presented above deals with the case $L+1=2$. It  
associates  to the three coordinates $x^i$ the three Pauli matrices, 
\ie, the three operators involved in the construction of 
$S_{fuzzy,2}$. With the identity matrix, they form the vector space 
of operators $M^2$.  We  introduced them  
through   their action on the Hilbert space  $V^{1/2}$ generated by 
$\Phi _1$ and $\Phi _2$, which  provides a 2-dimensional  IUR of 
SU(2). This   suggests 
the following generalisation of  the  CS procedure which leads to the 
fuzzy sphere $S_{fuzzy,L+1}$: we consider as   the Hilbert   space  Ê 
$V^{L/2}$  that of  the $(L+1)$-dimensional  IUR of SU(2). When 
$L=2k$ is even, we may chose for  $V^{k}$
 the canonical basis $\ket{k,i}$, $i=-k,...,k$, where each 
$\ket{k,i}\equiv Y^k_i$ is a spherical harmonic. 

This does not apply however when $L$ is odd. 
In the general case, we may follow \cite{novaes}. 
We select the  basis $\ket{L/2,i}$, $i=-L/2,...,L/2$,  corresponding 
to the  orthogonal functions $\Theta ^{L/2}_i$. These   functions are 
defined by the
 intermediary of the complex variable $z \equiv \tan \theta /2 
~e^{-i\phi}$, as
$\Theta ^{L/2}_i(x)=\Theta ^{L/2}_i(\theta,\phi)$\\ $ \equiv 
\sqrt{C^L_{L/2+i}}~\frac{z^{L/2+i}}{(1+\norm{z} 
^2)^{L/2}}=\sqrt{C^L_{L/2+i}}~\cos ^{L/2-i} \theta /2~\sin ^{L/2+i} 
\theta /2 ~~e^{-i(L/2+i)\phi}$, \\with $
,~C^L_{L/2+i} \equiv \frac{L!}{(L/2+i)!~(L/2-i)!}$ (formula (19) of 
\cite{novaes}).\\  This allows us to   see   $M^{L+1}$ as the set of 
endomorphisms End$(V^{L/2})$.

The coherent states are constructed following the procedure above:
$\ket{x}  = \sum _i \Theta ^{L/2}_i (x)~\ket{L/2,i}$. The observables 
are given by\begin{equation}
A_f = \sum _{i,j=-L/2}^{L/2} ~\int \mu(dx)~f(x)~ \overline{\Theta} 
^{L/2}_i (x)~ \Theta ^{L/2}_j  (x)~
\ket{j} \bra{i}.\end{equation}
In  other words, 
$[A_f]_{ij}=
\int \mu(dx)~f(x) ~ \overline{\Theta} ^{L/2}_i (x)~ \Theta ^{L/2}_j  
(x).$
Now we can develop $f$ as $f=\sum _{\ell=0}^{\infty} ~\sum _m 
~Êf_{\ell m} ~\Theta ^{\ell}_m$ and calculate the sum. To go further, 
we take into account the fact that the product $\overline{\Theta} 
^{L/2}_i (x)~ \Theta ^{L/2}_j  (x)$ can be developed themselves in 
spherical harmonics, with all terms having $\ell$ lower than $L$. 
Given the orthogonality of the spherical functions, this implies that 
the only terms in the development of $f$ are those with $\ell \le L$. 
Finally, this leads to \begin{equation}
F_{ij}=
\int \mu(dx)~\sum _{\ell =0}^L ~f_{\ell m} \Theta ^{\ell }_m(x) ~ 
\overline{\Theta} ^{L/2}_i (x)~ \Theta ^{L/2}_j  (x).
\end{equation} Involving the Clebsh-Gordan  
coefficients\begin{equation}
\label{ }
C^{\ell L/2 L/2}_{m i j} \equiv \int \mu(dx)~ \Theta ^{\ell }_m(x) ~ 
\overline{\Theta} ^{L/2}_i (x)~ \Theta ^{L/2}_j  (x), 
\end{equation}we obtain finally
\begin{equation}
F_{ij}=
 ~\sum _{\ell =0}^L ~f_{\ell m} ~C^{\ell L/2 L/2}_{m i 
j}.\end{equation}
In particular, the observables  $\hat{Y} ^{\ell }_m$ associated to 
the spherical harmonics $Y^{\ell }_m$, $\ell \le L$ are in number 
$(L+1)^2$ and provide a basis for $M^{L+1}$. 
They are defined by $[\hat{Y} ^{\ell }_m]_{ij}=C^{\ell L/2 L/2}_{m i 
j} .$

For any value of $L$, the CS construction leads to  an Hilbert space 
of  dimension $L+1$, as indicated above. What we have shown is that 
the canonical algebra of operators acting on $H$ identifies with the 
algebra of operators acting on $S_{fuzzy,L+1}$, the fuzzy sphere with 
$L+1$ cells. 

 \section{Discussion}

The CS quantization method proposed here applies to any Observation  
Set. In \cite{gghlr} we applied it to discrete samples and to the 
unit segment. More general developments will be given in a 
forthcoming paper. Here we have presented its application to the 
sphere $S^2$.  A quantization appear as a choice to look at the 
sphere with a different point of view, with a finite resolution. We 
have shown how complex quantizations lead to an  explicit 
construction of the Hilbert space associated  to the fuzzy sphere, 
although we have not examined the \nc ~differential structure 
associated. We have also emphasized the links with the theory of 
(irreducible) group representations.

The derivation of coherent spin states shows how this procedure, 
applied to a symplectic space, is able to give an usual dynamical 
quantum theory. However, as we claimed, it is much more general, 
allowing to perform quantization in the absence of any dynamical 
evolution. In further works, we will examine this possibility and 
study the application of  this quantization 
procedure to other manifolds, with and without  symplectic structure. 
Potential applications are the derivations of new fuzzy spaces. Also, 
since   quantization can be performed in the absence of any dynamics, 
this opens  perspective for fully covariant approaches, when no time 
function is defined.


\begin{thebibliography}{99}


\bibitem{csbook} Ali, S. T., Antoine, J.-P., and Gazeau, J.-P. : 
{\em Coherent states, wavelets and their generalizations}, 
{Graduate Texts in Contemporary Physics}, Springer-Verlag, New York, 
2000.

\bibitem{Arecchi}
Arecchi F.T. , Courtens  E. , Gilmore R. , and  Thomas, H., 
Phys. Rev. A, 6, 2211, 1972

\bibitem{ber} Berezin, F. A. : General concept of quantization, {\it 
Comm. Math. Phys.} {\bf 40} (1975), 153--174. 

\bibitem{daube} Daubechies, I. :
{\em Ten lectures on wavelets},
SIAM-CBMS, 1992.

\bibitem{csfks} Feng, D. H., Klauder, J. R., and Strayer, M. (eds.) : 
{\em Coherent States: Past, Present and Future (Proc. Oak Ridge 
1993)}, 
World Scientific, Singapore, 1994.

\bibitem{Freidel} Freidel L. and Krasnov K. 2002, Journal of Math. 
Phys. 43, 4 (april 2002)

\bibitem{gghlr} Gazeau, J-P., Garidi T., Huguet E., \lr ~M., Renaud 
J. 2003, 
Examples of Berezin-Toeplitz quantization: Finite sets and Unit 
Interval, 
to appear in proceedings of the Workshop in honor of R. Sharp, 
Montreal 2002, P. Winternitz ed., CRM-AMS

\bibitem{ga-kl} Gazeau, J-P., and Klauder, J. R. : Coherent states 
for systems with discrete and continuous spectrum, {\it J. Phys. A: 
Math. Gen. } {\bf 32} (1999), 123--132 . 

\bibitem{ga-mon} Gazeau, J-P., and Monceau, P. : Generalized coherent 
states for arbitrary quantum systems,
in {\it Conf\'erence Mosh\'e Flato 1999 -- Quantization, 
Deformations, and Symmetries\/}, edited by G. Dito and D. 
Sternheimer, Vol. II, pp. 131--144, Kluwer, Dordrecht, 2000. 

\bibitem{novaes} Gazeau J.-P. and Novaes 2003, J. Phys. A: Math. Gen. 
36 (2003) 199-212

\bibitem{klau1} Klauder, J. R. : Continuous-Representation Theory I. 
Postulates of continuous-representation theory, {\it J. Math. Phys. } 
{\bf 4} (1963), 1055--1058; Continuous-Representation Theory II.
Generalized relation between quantum and classical dynamics, {\it J. 
Math. Phys. } {\bf 4} (1963), 1058--1073. 


\bibitem{klau2} Klauder, J. R. : Coherent states for the hydrogen 
atom, {\it J. Phys. A: Math. Gen. }
{\bf 29} (1996), L293--298.

\bibitem{lali} Landau, L., and Lifshitz, E. M. : {\it Statistical 
physics}, Pergamon Press, 1958 

\bibitem{Elliott}  Lieb E. H. ,   Loss M., {\sl  Analysis
Graduate studies in Mathematics}, AMS, 2d Edition 2001


\bibitem{Madore} {\sl An Introduction to Noncommutative Differential 
Geometry and its Physical Applications}, Madore J. CUP  1995

\bibitem{magnus-ob} Magnus, W., Oberhettinger, F., and Soni, R. P. : 
{\sl Formulas and Theorems for the Special Functions of Mathematical 
Physics}, 3rd ed., Springer-Verlag, Berlin, Heidelberg and New York, 
1966.

\bibitem{Perelomov} Perelomov, A. M. : {\sl Generalized Coherent 
States and 
their 
Applications}, Springer-Verlag, Berlin, 1986. 

\bibitem{Radcliffe}
Radcliffe J.M. , {\sl Some properties of coherent spin states}, 
J. Phys. A : Gen. Phys., 4, 313, 1971


\end{thebibliography}
\end{document}